\documentclass[a4paper,12pt]{article}
\usepackage{amssymb,amsfonts,amsmath}
\usepackage{setspace}
\newtheorem{theorem}{Theorem}

\title{The inequality between mass and angular momentum for axially
  symmetric black holes}
\author{Sergio Dain\\
  Facultad de Matem\'atica, Astronom\'{i}a y F\'{i}sica\\
  Universidad Nacional de C\'ordoba\\
  Ciudad Universitaria \\
  (5000)  C\'ordoba\\
  Argentina.\\
  \\
  Max Planck Institute for Gravitational Physics\\
  (Albert Einstein Institute)\\
  Am M\"uhlenberg 1\\
  D-14476 Potsdam\\
  Germany.\\
email: dain@famaf.unc.edu.ar}

\begin{document}
\maketitle

\begin{abstract}
  In this essay I first discuss the physical relevance of the
  inequality $m\geq \sqrt{|J|}$ for axially symmetric (non-stationary)
  black holes, where $m$ is the mass and $J$ the angular momentum of
  the spacetime. Then, I  present a proof of this inequality
  for the case of one spinning black hole. The proof involves a
  remarkable characterization of the extreme Kerr black hole as an
  absolute minimum of the total mass. Finally, I conjecture on the
  physical implications of this characterization for the non linear
  stability problem for black holes.
\end{abstract}

\pagebreak
\doublespacing
\section{Introduction}

The Kerr metric is a solution of the vacuum Einstein equations which
depends on two parameters $m$ and $J$, the mass and the angular
momentum of the spacetime.  The metric is well defined for any choice
of the parameters, however, it only describes a black hole if the
following, remarkable, inequality holds
\begin{equation}
  \label{eq:1}
  m \geq \sqrt{|J|}.
\end{equation}
Roughly speaking, this inequality says that if an object is spinning too
fast it can not collapse to form a black hole.

Inequality \eqref{eq:1} has important consequences for a gravitational
collapse because the Kerr black hole is expected to play a unique role
in such process. To describe these consequences, let us
first review what is known as the standard picture of the
gravitational collapse. It mainly consists in the following two conjectures: i)
Gravitational collapse results in a black hole (weak cosmic
censorship). This conjecture point out the physical relevance of black
hole solutions. According, is it taken that solutions containing naked
singularity (for example, the Kerr solution whose parameters do not
satisfy \eqref{eq:1}) would not generically occur. ii) The
spacetime  settles down to a stationary final state, because
only a finite amount of gravitational radiation can be emitted by an
isolated system.  It is also reasonable to assume that at some finite
time all the matter fields have fallen into the black hole and hence
the exterior region is pure vacuum (for simplicity we discard
electromagnetic fields in the exterior). Then, the black hole
uniqueness theorem implies that the final state should be the Kerr
black hole. 

If the initial conditions for a collapse violate \eqref{eq:1} then
the extra angular momentum should be radiated away in gravitational
waves. There exists, however, an important class of spacetimes in
which angular momentum can not be radiated by gravitational waves:
axially symmetric spacetimes.  For an axially symmetric spacetime the
angular momentum is a conserved quantity. Then, the angular momentum
$J$ of the initial conditions must be equal to the final one $J_0$. On
the other hand, the mass of the initial conditions $m$ should be
bigger than the final mass of the resulting Kerr black hole $m_0$,
because gravitational radiation carries positive energy. If we assume
that i-ii) hold, then the system will settles down to a final Kerr
black holes, for which we have $m_0\geq \sqrt{|J_0|}$. Then, we deduce
that in this case the inequality \eqref{eq:1} should be satisfies by
the initial conditions. Also, if the initial conditions satisfy the
equality $m=\sqrt{|J|}$ no mass can be radiated and hence we expect
the system to be stationary.  Since the only stationary black hole
which satisfy this equality is extreme Kerr, in this case the system
should be exactly extreme Kerr. 

The argument presented above was essentially given in
\cite{Friedman82} and it is similar to the one used by Penrose
\cite{Penrose69} to obtain the inequality between mass and the area of
the horizon on the initial data. As in the case of Penrose inequality,
a counter example of \eqref{eq:1} will imply that the standard picture
of the gravitational collapse is not true. Conversely a proof of
\eqref{eq:1} gives indirect evidence of its validity, since it is very
hard to understand why this highly nontrivial inequality should hold
unless i)-ii) can be thought of as providing the underlying physical
reason behind it (see the discussion in \cite{Wald99} \cite{Penrose98}).

The physical interpretation of \eqref{eq:1} in the non stationary case
is the following. If we
have a stationary vacuum black hole (i.e. Kerr) with mass $m_0$ and
angular momentum $J_0$ and add to it axisymmetric gravitational waves,
then the spacetime will still contain a (non-stationary) black hole,
these waves will only increase the mass and not the angular momentum
of the spacetime because they are axially symmetric.  Since Kerr
satisfies $m_0\geq \sqrt{|J_0|}$, then we get \eqref{eq:1} for the
resulting spacetime. The difference $m-\sqrt{|J|}$, which can be
calculated a priori on the initial conditions and by \eqref{eq:1} is
positive, provides an upper bound for the total amount of radiation
$E$ emitted by the system
\begin{equation}
  \label{eq:2}
E=m-m_0 \leq  m-\sqrt{|J_0|}= m-\sqrt{|J|}.
\end{equation} 
This is the same  argument as the one used by
Hawking in \cite{Hawking72}.

\section{The variational approach}
Inequality \eqref{eq:1} suggests a variational principle, namely,
extreme Kerr realizes the minimum of the mass among all axially symmetric
black holes with fixed angular momentum. However, it is important to
note that for two related inequalities, the positive mass theorem and the Penrose
inequality, a variational formulation was not successful. In
the case of the positive mass theorem only a local version was proved
using a variational principle \cite{Choquet-Bruhat76}.   

The key difference in the present case is axial symmetry. In that case
it possible to write the mass (in an appropriate gauge) as a positive
definite integral on a spacelike hypersurface.  The reason for this
particular behavior of the mass is the following. In the presence of a
symmetry, vacuum Einstein equations can be reduced a la Kaluza-Klein
to a system on a 3-dimensional manifold where it takes the form of
3-dimensional Einstein equations coupled to a matter source.  Since in
3-dimension there is no radiation (the Weyl tensor is zero), this source
represents the true gravitational degree of freedom that have
descended from 4-dimensions to appear as ``matter'' in 3-dimension.
Since all  the energy is produced by these effective
matter sources,  one would expect in that,  as in other
field theories, the total energy of the system can be expressed as a
positive definite integral over them. This was in fact proved
by Brill \cite{Brill59} in some restricted cases and then generalized
in \cite{Gibbons06} \cite{Dain06c}.

The mass integral essentially depends on two free functions: the norm
and the twist potential of the axial Killing vector. This allow as to
make unconstrained variations of them and hence formulate a well
defined  variational problem for the mass functional.

In a series of recent articles \cite{Dain05c}\cite{Dain05d}
\cite{Dain05e} \cite{Dain06c}, the inequality \eqref{eq:1} has been
proved for the case of one spinning black holes using this variational
formulation. These results can be summarized as follows (we have
suppressed some technical assumptions, for a precise
formulation see \cite{Dain06c}). 
\begin{theorem}
\label{t:1}
Extreme Kerr realizes  the unique absolute minimum of the mass functional among all
axially symmetric spinning black holes with only one connected component and fixed
angular momentum.     
\end{theorem}
While restricted to axially symmetric solutions, this theorem
represent the first non linear and non stationary result concerning
the Kerr black hole.

Black holes are critical points of the mass at fixed angular momentum
and horizon area \cite{Sudarsky91}. This is, in essence, the first law
of black hole mechanics.  However, only extreme Kerr is in addition a
minimum of the mass.

This characterization of extreme Kerr as a minimum of the mass implies
a kind of stability of this solution. It is a priori not directly
related to the non linear stability of extreme Kerr under the
evolution of Einstein equations. But it suggests that the extreme Kerr
black hole is stable in a more fundamental way than Schwarzschild or
non-extreme Kerr among axially symmetric deformations.

This suggestion is supported by quantum effects.
They are two quantum effects which can make the black hole unstable:
Hawking radiation and particle production by superradiance. For the
extreme Kerr black hole the temperature is zero and hence there is no
Hawking radiation. The superradiance effect is related to the transfer
of angular momentum from the black hole to the exterior (similar to
the Penrose process).  If we restrict ourself to axially symmetric
configurations, this transfer is not possible. In this sense,
extreme Kerr black hole is quantum stable among axially symmetric
configurations.

The non linear stability of black holes is a major open problem in
General Relativity. The first non trivial vacuum model to be studied
is represented by axisymmetric spacetimes.  The results presented here
reveal the following two relevant features of axial symmetry, which
are likely to play an important role in this problem.  First, the mass
is a positive definite integral on the spacelike hypersurfaces.  Since
the mass is a conserved quantity, the norm defined by this  integral 
controls the fields during the evolution.  Second, the above
considerations indicate that in the class of axisymmetric solutions
the extreme Kerr black hole possesses  hidden properties
which are not present in the non extreme case. Their analysis may
significantly simplify the study of its non linear stability.

\singlespacing

\section*{Acknowledgments}
The author is supported by CONICET (Argentina).
This work was supported in part by grant PIP 6354/05 of CONICET
(Argentina), grant 05/B270 of Secyt-UNC (Argentina) and the Partner
Group grant of the  Max Planck Institute for Gravitational Physics,
Albert-Einstein-Institute (Germany). 

I would like to thank Helmut Friedrich for reading the manuscript.
I would like also to thank the hospitality and support of the Max
Planck Institute for Gravitational Physics (Albert Einstein Institute)
were most of the writing of this essay  took place.


\end{document}